# Video-rate mid-infrared imaging in the molecular fingerprint region via nanosecond non-degenerate two-photon absorption

*Evan P. Garcia[1], Yryx Y. L. Palacios[2], Ryan Nguyen[1], Adam Hanninen[3], Aleksei I. Noskov[1], Eric O. Potma[1*] and Dmitry A. Fishman[1*]*

[1]*Department of Chemistry, University of California Irvine, Irvine, CA 92697*
[2]*Beckman Laser Institute and Medical Clinic, University of California Irvine, Irvine, CA 92697*
[3]*Trestle Optics, Irvine, CA 92697, USA*

dmitryf@uci.edu
epotma@uci.edu

### Abstract

Non-degenerate two-photon absorption (NTA) offers an attractive route for wide-field mid-infrared (MIR) imaging by mapping long wavelength information into the spectral detection windows of mature near-infrared detector technologies. However, existing NTA implementations rely almost exclusively on complex, large-footprint femtosecond laser systems, severely limiting practicality and scalability. Here, we demonstrate an NTA imaging platform that replaces the ultrafast laser with a compact nanosecond mid-IR source coupled to a high-definition indium gallium arsenide (InGaAs) camera. Operating in the nanosecond regime removes stringent temporal-overlap requirements, dramatically simplifying system architecture while preserving high nonlinear sensitivity. Using this approach, we achieve chemically selective, wide-field imaging deep into the mid-IR molecular fingerprint region and demonstrate, for the first time, video-rate NTA imaging in this spectrally rich regime. By combining relaxed alignment constraints, compact excitation, and high-speed fingerprint-region imaging, this work establishes nanosecond NTA as a practical and scalable foundation for next-generation mid-IR chemical imaging.

INTRODUCTION

The mid-infrared (mid-IR) spectral region (3–12 µm) hosts a rich set of fundamental vibrational resonances that provide intrinsic chemical contrast for applications spanning medicine [1,2], biology [3], manufacturing [4,5], and defense [6]. Despite its importance, practical mid-IR imaging remains limited by the trade-offs associated with existing detection strategies. Direct detection with mid-IR cameras offers broad spectral coverage, but is constrained by cost, limited pixel numbers, or low frame rates [7,8]. Furthermore, the small bandgaps of mid-IR detector materials result in high thermal noise, often necessitating cryogenic cooling to achieve acceptable performance [9,10].

Recent approaches seek to circumvent the fundamental challenges of mid-IR detection and imaging by mapping mid-IR spectral information into the near-infrared or visible domains, where detection technologies are significantly more efficient, mature, and readily available [11-20]. One such approach exploits non-degenerate two-photon absorption (NTA), in which a mid-IR photon is combined in time and space with a higher-energy visible or near-infrared photon to directly induce an interband optical transition in the semiconductor sensor material [13,21]. This mechanism enables the use of wide-bandgap semiconductor detectors and focal plane arrays based on well-established platforms such as silicon (Si) and indium gallium arsenide (InGaAs), which offer superior performance compared to conventional mid-IR detectors, particularly in terms of reduced thermal noise, higher pixel density, and operational stability. By leveraging these mature detection technologies, NTA-based mid-IR imaging overcomes many of the intrinsic limitations of traditional mid-IR cameras and focal plane arrays [22,23]. Importantly, as a $\chi^3$-mediated nonlinear absorption process, NTA does not require phase matching, it accommodates an ultrabroad spectral range across the near-to-far IR, and it does not rely on complex image reconstruction. These attributes significantly simplify implementation and enhance robustness against environmental drift and external perturbations. Leveraging these advantages, NTA-based mid-IR (MIR) imaging modalities have enabled high-speed chemically selective high-definition imaging [22,24], high-speed hyperspectral imaging [25], and volumetric chemically selective imaging [26].

Despite addressing several fundamental challenges in mid-IR imaging, detectable NTA signals generated in the thin semiconductor layers (<30 µm dwell thickness) of modern camera sensors require optical irradiances substantially higher than those used in linear detection schemes. As a result, most existing demonstrations rely on ultrafast femtosecond or picosecond laser sources, whose large footprint, high cost, and operational complexity limit the applicability of NTA imaging outside specialized optical laboratories [23,25-28]. Advancing NTA toward practical and deployable MIR imaging systems, therefore, necessitates the development of more compact and accessible excitation sources.

In this work, we address the complexity of light sources typically required for NTA by demonstrating an imaging platform based on a compact nanosecond light source that provides both the MIR excitation and the near-infrared gate pulse. This approach represents a critical step toward bridging the gap between proof-of-principle demonstrations and practical applications. In addition to lower cost and reduced footprint, a key advantage of nanosecond excitation over femtosecond sources is the relaxation of maintaining the sensitive temporal overlap between the mid-IR and gate pulses. Because nanosecond pulses are orders of magnitude longer in duration, temporal overlap is readily maintained over standard optical path lengths, significantly simplifying experimental implementation. Using a compact optical parametric oscillator (OPO) pumped by a nanosecond laser, we achieve high-speed, wide-field mid-IR imaging across the MIR molecular fingerprint region. We demonstrate chemically specific contrast in both macroscopic and microscopic imaging modes, including real-time, video-rate analysis of bio-relevant materials. These results establish that more affordable nanosecond excitation can efficiently drive NTA-based detection in the mid-IR fingerprint region using compact sources and standard detectors, providing a viable pathway toward practical, scalable MIR imaging systems.

MATERIALS AND METHODS

Sample Preparation

Polymer samples were melted, deposited onto 5 mm-thick NaCl windows, and pressed flat using a $CaF_2$ coverslip. L-proline crystals were mounted on IR-transparent KBr windows. Type I collagen from rat tail tendon was flattened, mounted on $CaF_2$ microscope slides, and sealed with a $CaF_2$ coverslip. Tardigrades were pipetted from an aqueous suspension, allowed to dry under ambient conditions, and sealed beneath a $CaF_2$ coverslip.

FTIR Analysis

Infrared absorption spectra were acquired using a Jasco 4700 FTIR spectrometer equipped with a diamond-based attenuated total reflection (ATR) accessory. Spectra were collected with a 2 cm$^{-1}$ resolution and averaged over 16 scans. Tardigrade spectra were obtained with a Nicolet iS20 FTIR spectrometer coupled to a Thermo Scientific Nicolet Continuum microscope using a confocal Reflachromat 15×, 0.58 NA objective. Data were collected with a resolution of 2 cm$^{-1}$ and averaged over 16 scans.

Tunable nanosecond light source

For ns-NTA excitation, we developed a zinc germanium phosphide ($ZnGeP_2$) optical parametric oscillator (OPO) configured for type-I phase matching. $ZnGeP_2$ was selected for its large nonlinear coefficient (~75 pm/V) and favorable thermal properties, enabling broad, continuous mid-infrared tunability spanning 2.5 –10 µm, encompassing both the signal and idler beams [29,30]. The OPO was pumped by a Q-switched Ho:YAG laser (2090 nm, 45 ns pulses, 1 kHz, 10 W average power; HLPN 2090, IPG) and implemented in a compact X-shaped confocal resonator ring cavity (L=60 cm), resonant with the signal beam, with two concave intra-cavity mirrors (radius of curvature R=100 mm) and a partially transmitting output coupler [31].

NTA setup

A schematic of the experimental setup is shown in Fig. 1. The idler output of the OPO, separated from the residual pump using a custom dichroic mirror (2090 nm cutoff), was pre-calibrated and spectrally characterized over the 5–7 µm range and served as the MIR light source in the imaging experiments. The residual 2.09 μm OPO pump radiation was spatially and temporally overlapped with the tunable MIR idler beam to generate a non-degenerate two-photon absorption signal directly on the InGaAs camera chip (bandgap 0.73 eV, 1280 × 1020 pixels, 12 μm pixel pitch; MVCam, Princeton Infrared Technologies). The MIR beam was delivered through an imaging system composed of either a 2x magnification $CaF_2$ telescope or an unobscured all-reflective objective (NA = 0.65) paired with a $CaF_2$ tube lens (f = 150 mm). The resulting pulse irradiance at the camera chip was set to 2.67 kW cm$^{-2}$ for the MIR beam and 4.45 kW cm$^{-2}$ for the 2.09 μm gate radiation.

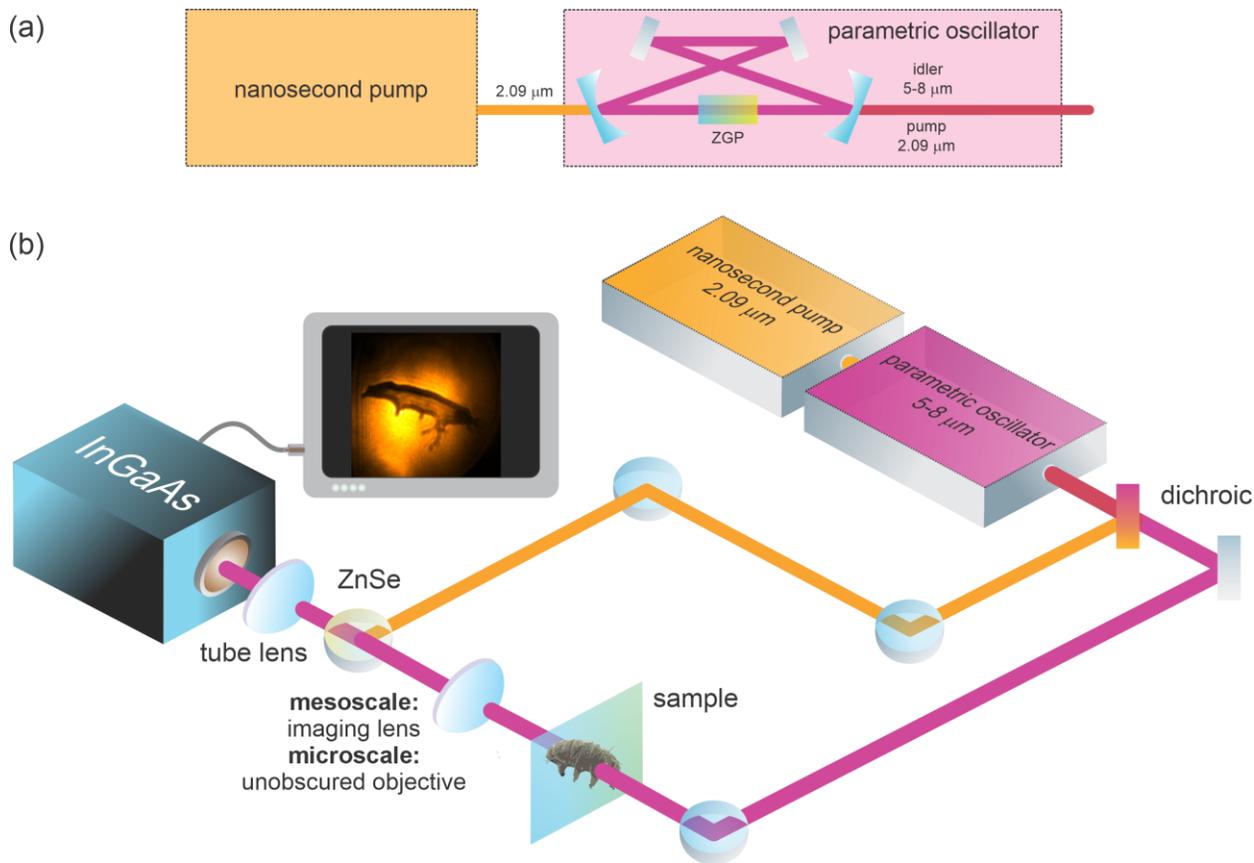

**Figure 1.** (a) Schematic of a nanosecond Ho:YAG laser (2.09 μm) seeding a compact $ZnGeP_2$ (ZGP)–based optical parametric oscillator (OPO) to generate tunable mid-infrared idler pulses spanning the molecular fingerprint region (≈60 ns pulse duration, 10 W, 1 kHz repetition rate). While the OPO design supports broad tunability from approximately 2.6 to 10 μm, the experiments reported here were performed in the pre-characterized 5–8 μm spectral range. (b) Nanosecond non-degenerate two-photon absorption (ns-NTA) imaging setup, in which the tunable mid-IR idler serves as the excitation beam and the residual 2.09 μm pump pulse acts as the gate pulse. The two beams are spatially and temporally recombined on an InGaAs camera (MVCam, Princeton Infrared Technologies Inc.; 1280 × 1020 pixels, 12 μm pixel pitch, 0.73 eV bandgap). Two imaging configurations were employed: mesoscale imaging using 75 mm $CaF_2$ lens (NA = 0.015) and microscale imaging using a custom-designed, unobscured reflective objective (NA=0.65) [32].

## Results

*Static mesoscale imaging in the molecular fingerprint range*

To demonstrate the potential of nanosecond NTA (ns-NTA) imaging in the molecular fingerprint region and to achieve label-free, chemically selective contrast, we examined a variety of organic samples. For these mesoscale imaging experiments, a two-times expansion of the MIR beam was used to match the beam size to the camera chip area (~1.9 cm²). As an initial demonstration, we investigated a hybrid sample consisting of two adjacent thin polymer films: polylactic acid (PLA) and polyvinyl chloride (PVC). Figure 2a shows the FTIR spectra of PLA and PVC, with specific spectral points marked, where ns-NTA imaging was performed (Figures 2b–2g). Consistent with the infrared absorption spectra of each polymer, the NTA images reveal clear chemical contrast at selected frequencies. For example, the PVC material appears dark due to strong absorption at 1420 cm$^{-1}$ (Figure 2b), whereas the transmission contrast is reversed at 1760 cm$^{-1}$ owing to strong absorption by PLA (Figure 2f). All NTA images in Figure 2 were normalized and corrected for small variations in excitation power.

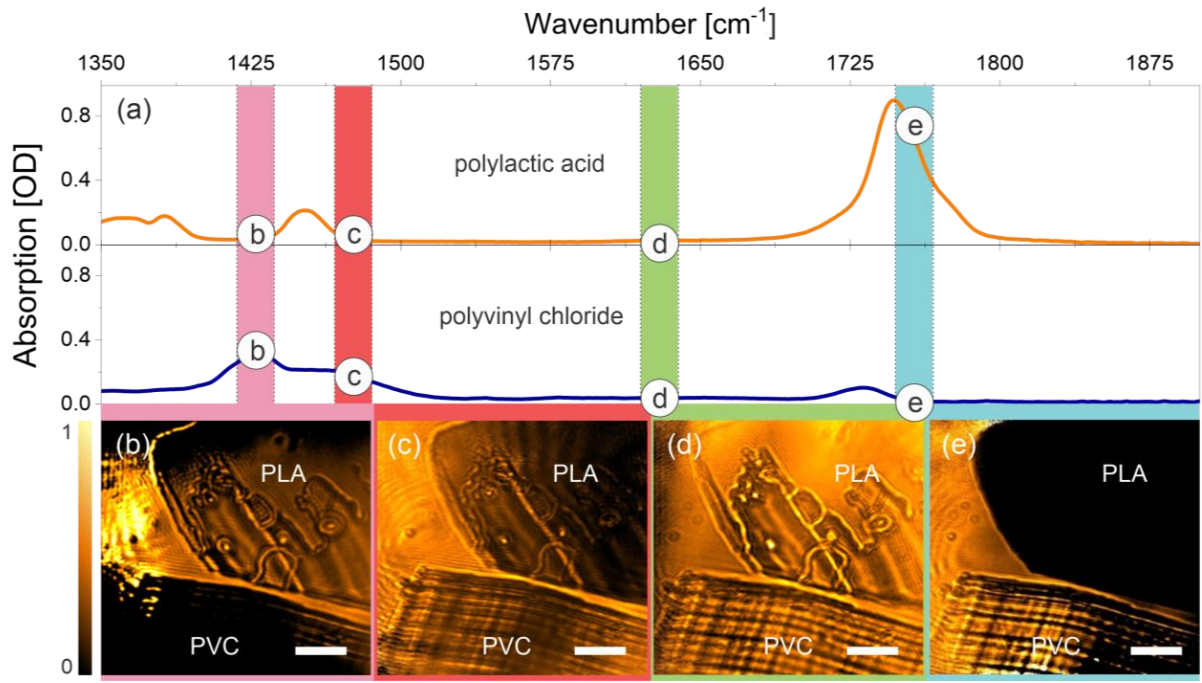

**Figure 2.** Static mesoscale ns-NTA imaging. (a) FTIR absorption spectra of polylactic acid (PLA, orange) and polyvinyl chloride (PVC, blue) polymers joined in a single hybrid sample. (b–e) Wide-field mid-IR images acquired at selected spectral points indicated in (a). The images show strong absorption contrast consistent with the characteristic vibrational features of each polymer. Scale bar: 1 mm. Exposure time: 30 ms. All images are normalized and corrected for small variations in excitation power.

*Dynamic imaging in the fingerprint range*

Further demonstration of the ns-NTA capability was performed using L-proline, a common amino acid involved in key cellular and metabolic processes and essential for protein synthesis. The absorption spectrum of crystalline L-proline is shown in Figure 3a, highlighting a characteristic resonance associated with the carbonyl stretching mode at 1560 cm$^{-1}$. Following the absorption spectrum, Figure 3 presents static mesoscale ns-NTA images of crystalline L-proline clusters acquired on-resonance (1560 cm$^{-1}$, Figure 3b) and off-resonance (1800 cm$^{-1}$, Figure 3c). To demonstrate chemically selective dynamic imaging in the fingerprint region, we recorded a video of L-proline crystals dissolving in isopropanol alcohol (IPA). The video shows that the crystalline amino-acid clusters fully dissolve within only a few seconds, while simultaneously revealing both the evolving cluster morphology and a pronounced change in absorption contrast as solvation alters the local L-proline concentration. This high-definition (~1.2 Mpx) video was recorded at ~33 frames per second (30 ms per frame) and is presented in Supplementary Video 1.

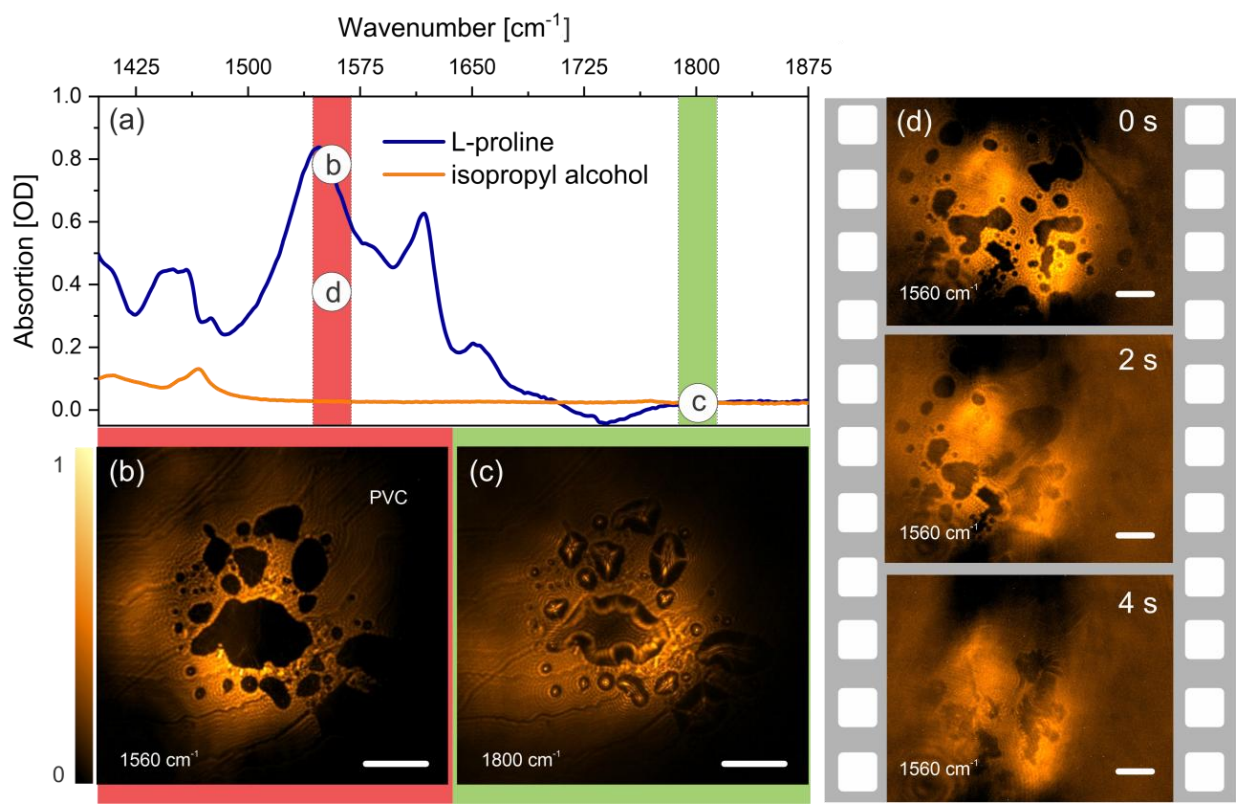

**Figure 3.** Dynamic imaging in mid-IR fingerprint region. (a) FTIR absorption spectra of L-proline (blue) and isopropyl alcohol (IPA, orange). (b,c) Representative wide-field ns-NTA images acquired on-resonance at the carbonyl stretching band (1560 cm$^{-1}$) and off-resonance (1800 cm$^{-1}$), respectively, showing clear absorption contrast. Scale bar 1 mm. Exposure time 30 ms. (d) Snapshot from a time-resolved ns-NTA video visualizing the dissolution of L-proline clusters in real time, recorded at 33 frames/s at the carbonyl stretching resonance. The video reveals cluster dissolution through concurrent changes in morphology and local absorption contrast.

*Microscale imaging of collagen and microorganisms*

To extend NTA-based mid-IR imaging to the microscale, we employed a setup incorporating a custom-designed, unobscured, all-reflective microscope objective (Figure 4a). The objective design and its characterization for both linear and nonlinear imaging have been described previously [32]. As representative microscale samples, we investigated rat tail collagen tissue and a tardigrade (*Milnesium tardigradum*), a microscopic organism commonly found in aqueous environments. Figure 4a shows the FTIR spectra of both samples, highlighting the characteristic amide I absorption band near 1650 cm$^{-1}$.

Figures 4b,c (rat tail collagen) and 4d,e (tardigrade) present ns-NTA images acquired on- and off-resonance using two spectral points: the amide I resonance at 1650 cm$^{-1}$ and an off-resonant reference at 1800 cm$^{-1}$. Figures 4f and 4g show the corresponding composite images, in which regions exhibiting resonant absorption are rendered in red, while spectrally transparent regions appear yellow, forming false-color mid-IR images of the biological samples. All images exhibit clear and well-defined absorption contrast with acquisition times as short as 30 ms, demonstrating that ns-NTA can be extended to microscopic length scales while preserving both chemical specificity in the molecular fingerprint region and high imaging speed. We note that the ns-NTA fingerprint image of the tardigrade was acquired on a dehydrated (non-living) specimen. Although tardigrades are exceptionally resilient—capable of surviving extreme environments, including exposure to outer space—maintaining a living and motile organism requires an aqueous environment. For relatively large and optically thick living specimens, strong mid-IR absorption by water fundamentally limits photon transmission in the fingerprint region, imposing practical constraints on chemically specific imaging under hydrated conditions. Under dehydrated conditions, however, ns-NTA imaging at 1650 cm$^{-1}$ yields clear vibrational contrast associated with protein-rich structures, with acquisition times comparable to those used for collagen (Figures 4d and 4g).

Taken together, these results establish that ns-NTA enables chemically specific mid-IR imaging at microscopic length scales within the molecular fingerprint region, while maintaining short acquisition times compatible with dynamic measurements. This capability extends NTA beyond previously demonstrated mesoscale implementations and confirms that nanosecond-driven excitation can support high-speed, label-free vibrational imaging of biological specimens with microscale spatial resolution.

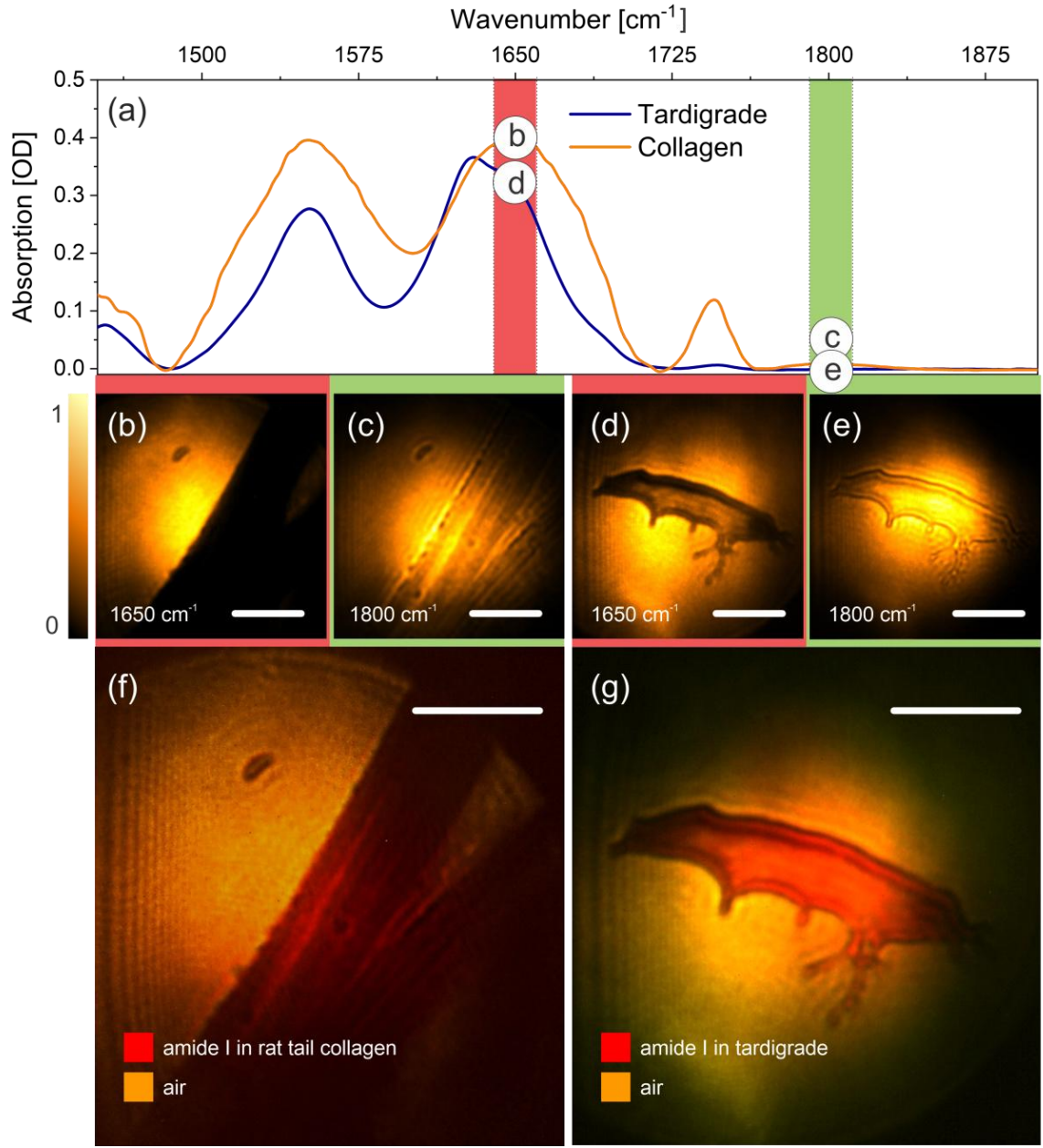

**Figure 4.** (a) FTIR absorption spectra of rat tail collagen (blue) and dehydrated tardigrade (*Milnesium tardigradum*, orange). Both spectra exhibit a pronounced absorption feature near 1650 cm$^{-1}$, corresponding to the amide I vibrational band. (b–e) ns-NTA images acquired on- and off-resonance of the amide I band for collagen (b,c) and tardigrade (d,e). Scale bar 100 μm. Exposure time 30 ms. (f,g) Corresponding false-color composite images constructed from the on- and off-resonance image pairs for collagen (f) and tardigrade (g). Scale bar 100 μm. Regions exhibiting amide I absorption are rendered in red, while areas with negligible absorption appear yellow.

**Discussion**

The results presented here reposition non-degenerate two-photon absorption (NTA) as a practical and scalable strategy for mid-infrared imaging, rather than a technique confined to ultrafast optics laboratories. By demonstrating NTA driven entirely by nanosecond excitation, this work reduces the complexity, cost, and footprint of NTA-based imaging systems, while also enabling portability and mobility. In contrast to previous implementations relying on femtosecond or picosecond laser sources [27,33,34], the nanosecond platform presented here is readily compatible with standard laboratory environments and is intrinsically amenable to miniaturization. Ongoing advances in compact nanosecond light sources suggest a clear pathway toward future field-deployable implementations. This transition directly addresses one of the central bottlenecks that has limited the broader adoption of NTA imaging despite its intrinsic advantages.

A key conceptual advantage of nanosecond excitation lies in the relaxation of otherwise stringent temporal-overlap requirements. In ultrafast NTA implementations, femtosecond pulse durations necessitate precise delay-line control and continuous optimization to maintain temporal overlap between the mid-IR and gate pulses. In the nanosecond regime, pulse durations are sufficiently long that temporal overlap is effectively guaranteed once spatial overlap is achieved. As a result, NTA becomes a geometrically rather than temporally constrained process. This dramatically simplifies system alignment, enhances long-term stability, and renders the optical layout highly forgiving to environmental drift. Importantly, this relaxed timing constraint opens pathways that are impractical for ultrafast systems, including the use of optical fibers, flexible beam delivery, and compact or modular architectures, all of which are essential for deployable and user-friendly mid-IR imaging platforms.

An important implication of operating in the nanosecond regime is the expanded compatibility with emerging mid-IR light sources that are incompatible with ultrafast NTA implementations. In particular, quantum cascade lasers (QCLs) offer compact form factors, intrinsic mid-IR emission, and rapid electronic tunability that is well-suited for hyperspectral imaging. While current commercially available QCL sources generally do not yet provide sufficient peak irradiance to reliably drive NTA in standard camera sensors, the relaxed temporal constraints of ns-NTA make QCL integration a realistic long-term pathway rather than a conceptual mismatch. Continued advances in high-peak-power QCL designs, along with detector architectures optimized for nonlinear absorption, could enable solid-state, electrically driven NTA imaging platforms with rapid wavelength agility.

Crucially, the gains of ns-NTA simplicity do not come at the expense of performance. On the contrary, nanosecond-driven NTA demonstrated here achieves chemically specific imaging deep into the molecular fingerprint region, representing the first realization of NTA imaging in this spectrally rich regime. This result highlights a defining strength of NTA as a $\chi(3)$-mediated process: its intrinsic lack of phase-matching constraints enables seamless and *gapless spectral coverage extending from the near-infrared into the mid- and, in principle, far-infrared using a single camera chip* [35]. Furthermore, the increase in photon energy disparity enhances the NTA response [36], directly benefiting operation in the fingerprint region, where vibrational resonances are strongest. This behavior underscores the fundamental scalability of NTA and distinguishes it from alternative frequency-conversion approaches that rely on separate nonlinear crystals and strict phase-matching conditions [37,38].

Beyond spectral reach, the ns-NTA platform enables high-speed, wide-field imaging with acquisition times short enough to exceed standard video rates. This capability is particularly significant given that fingerprint-region imaging is traditionally associated either with slow point-scanning modalities or with limited availability of high-definition focal plane arrays sensitive in this spectral range. While performance in aqueous environments remains constrained by mid-IR attenuation in water, the demonstrated acquisition speeds establish that, where optical transmission permits, wide-field nanosecond NTA allows chemically selective imaging with millisecond-scale frame times, enabling the observation of dynamic chemical and physical processes in real time.

**Conclusions**
This work establishes nanosecond-driven non-degenerate two-photon absorption as a practically distinct and scalable implementation of mid-infrared imaging. By eliminating ultrafast timing constraints, ns-NTA decouples chemical specificity and spectral reach from system complexity, enabling compact, robust, and intrinsically mobile imaging architectures without sacrificing performance. The demonstration of wide-field, video-rate imaging in the molecular fingerprint region highlights the capability of ns-NTA to access vibrational contrast with substantially greater simplicity than previous NTA implementations, while preserving performance in terms of sensitivity, spatial resolution, and acquisition speed. Together, these results define a new design space for NTA-based imaging systems and provide a clear pathway toward deployable mid-IR chemical imaging technologies.

**Acknowledgements**
D.A.F. thanks Yulia Davydova for support and help during this study. D.A.F. and E.O.P acknowledge funding from Chan Zuckerberg Initiative 2023-321174 (5022) GB-1585590, NSF 2025-2434622. A.H. acknowledges NIH R43GM149004.

**Declaration of competing interest**
A.H. declares a financial interest in Trestle Optics. All other authors declare no competing interests.